\documentclass[11pt]{article}

\usepackage[utf8]{inputenc}
\usepackage[T1]{fontenc}
\usepackage{lmodern}
\usepackage{microtype}
\usepackage[margin=1in]{geometry}
\usepackage{graphicx}
\usepackage{booktabs}
\usepackage{array}
\usepackage{amsmath}
\usepackage{amssymb}
\usepackage[hidelinks]{hyperref}
\usepackage{caption}
\title{RAGAL: A Frugal, Fully Local Retrieval-Augmented Assistant\\for Technical Support at a Government Agency}
\author{Dan Mu\textcommabelow{s}etoiu\\
IT Department, Agency for Financing Rural Investments (AFIR), Romania\\
\texttt{dan.musetoiu@afir.info}}
\date{July 2026}

\begin{document}

\maketitle

\begin{abstract}
Public institutions hold large volumes of sensitive documents and support
tickets that cannot leave the premises, ruling out cloud-hosted language
models entirely. We report on RAGAL, a retrieval-augmented assistant for the
technical-support team of AFIR, the Romanian Agency for Financing Rural
Investments, built and operated under three hard constraints: zero data
egress (no external API calls, even for synthetic data), a read-only mandate
(the assistant drafts, humans execute), and a single 8~GB consumer laptop as
the only development and training machine. Over a Romanian-language corpus of ${\sim}25{,}000$ chunks ---
15{,}073 resolved support tickets and internal normative documents --- we
show that the highest-leverage investments were retrieval engineering and
retriever fine-tuning rather than a larger generator: hybrid dense--sparse
retrieval with intent routing raised our internal evaluation from 62\% to
81\%, and fine-tuning the bge-m3 embedder on real ticket data improved
recall@10 from 0.663 to 0.850 (MRR $0.489 \rightarrow 0.684$) after 72
minutes of training. We document a general pitfall: single-domain fine-tuning
silently degraded retrieval on the untouched document domain below the stock
baseline, detected only after building a per-domain evaluation set and
repaired with locally generated queries (GenQ). We report two counter-intuitive
findings --- PII masking \emph{improved} generation quality, and a
structural ``anchor distillation'' scheme made SQL hallucination impossible
by construction --- along with a reproducible recipe for full embedder
fine-tuning in 8~GB of VRAM. Finally, since zero egress also rules out a cloud
judge, we describe a substitute: a 744B-parameter model run on CPU, too slow
to serve interactively but affordable in overnight batch, used as a second
opinion whose limits we quantify. We release the sanitized pipeline scripts
for institutions facing similar data-locality constraints.
\end{abstract}

\noindent\textbf{Keywords:} retrieval-augmented generation, on-premise LLM,
embedder fine-tuning, low-resource deployment, Romanian NLP, public
administration

\section{Introduction}

Large language models have made assistant-style tooling accessible to almost
any organization --- provided the organization can send its data to a cloud
API. For many public institutions this is not an option. The corpus that
would make an assistant useful (internal procedures, support tickets
containing personal data of beneficiaries, database corrections) is exactly
the corpus that must never leave the building.

This paper is an experience report on building such an assistant anyway,
under constraints that we believe are representative for the public sector:

\begin{itemize}
\item \textbf{C1 --- Zero egress.} No token of the corpus may reach an
external service, including for auxiliary tasks such as synthetic
training-data generation. Everything --- generation, embedding, fine-tuning,
evaluation --- runs on hardware inside the institution.
\item \textbf{C2 --- Read-only mandate.} The assistant suggests; humans
execute. It drafts SQL correction scripts anchored in past precedents, but
has no write access to any system, structurally eliminating the risk of
automated damage.
\item \textbf{C3 --- Frugality.} The entire system was developed, trained
and piloted on one consumer laptop (NVIDIA RTX 2000 Ada, 8~GB VRAM, 32~GB
RAM), with no dedicated inference server during development.
\end{itemize}

RAGAL (``RAG for GAL'', after the Local Action Groups --- \emph{Grupuri de
Ac\textcommabelow{t}iune Local\u{a}} --- whose activity the agency
coordinates) supports the team that maintains the agency's application
ecosystem: it answers questions about procedures and application workflows,
retrieves precedents from 15{,}073 resolved support tickets, and drafts
parameterized SQL correction scripts for a technician to review and run.

Our contributions are practical rather than algorithmic:

\begin{enumerate}
\item An end-to-end account of a fully local, Romanian-language RAG system
in pilot production over a dual corpus (normative documents + support
tickets), including the engineering decisions that mattered most.
\item A frugal recipe for \emph{full} fine-tuning of a 568M-parameter
embedder in 8~GB of VRAM (8-bit optimizer + gradient checkpointing), with a
diagnostic for the silent VRAM-spill failure mode on Windows consumer
hardware. Training took 72 minutes and improved recall@10 by 19 points.
\item An empirically documented pitfall: fine-tuning the retriever on one
domain (tickets) silently \emph{regressed} the other domain (documents)
below the stock model --- invisible until we built a per-domain evaluation
set --- and its zero-egress repair via locally generated synthetic queries.
\item Two counter-intuitive observations: (a)~masking personally
identifiable information (PII) in the corpus \emph{improved} downstream
generation quality; (b)~letting the LLM write only prose around
verbatim-copied SQL (``anchor distillation'') eliminates SQL hallucination
structurally.
\item Negative results we found valuable: small vision LLMs are not usable
as OCR; community ``reasoning/uncensored'' model variants confabulated legal
thresholds under explicit abstention instructions; and three newer embedding
models failed to beat a well-chosen 2024 baseline on our corpus.
\item A zero-egress substitute for LLM-as-a-judge: a 744B-parameter MoE model
running on CPU with experts streamed from NVMe, unusable interactively at
0.2--0.3~tok/s but affordable in overnight batch because prefill scales
sub-linearly. We report what it costs, what it found, and --- from a human
meta-audit of a third of its verdicts --- how far its individual judgments
can be trusted.
\end{enumerate}

\section{Related Work}

RAG remains the standard architecture for grounding LLMs in private corpora
\cite{lewis2020rag}. For embedder adaptation we build on
sentence-transformers \cite{reimers2019sbert}, the multiple-negatives
ranking loss \cite{henderson2017efficient}, hard-negative mining, and
GenQ/GPL-style synthetic query generation \cite{wang2022gpl}. Our decision
to fine-tune the retriever rather than the generator is consistent with
recent evidence that language-specific LLM fine-tuning does not reliably
beat strong multilingual bases --- e.g., on the Romanian GRILE benchmark,
base gemma-2-9b-it outperformed all RoGemma2 fine-tunes
\cite{dumitran2025grile}. Finetune-RAG \cite{lee2025finetunerag} proposes
SFT against fabricated context for factual discipline; we identify it as
future work for our generator. Unlike benchmark-oriented work, we report a
deployed system under explicit data-locality constraints, in the spirit of
industry-track experience reports.

\section{Setting and Constraints}

AFIR administers EU rural-development funds; the DR-36 LEADER intervention
is implemented by 246 Local Action Groups (GALs) through a dedicated
application portal. The support team behind this portal --- analysts,
developers, and domain experts --- resolves tickets ranging from account
issues to workflow deadlocks and database corrections.

The corpus spans two registers --- formal normative prose and colloquial
ticket language --- but a single language: Romanian, with inconsistent
diacritics (the same word may appear with and without diacritical marks
across tickets), which turns out to matter for embedding stability
(\S6.1).

Table~\ref{tab:constraints} summarizes the constraint-to-decision mapping
that shaped the system.

\begin{table}[htbp]
\centering
\small
\caption{Constraints and the design decisions they forced.}
\label{tab:constraints}
\begin{tabular}{p{0.30\linewidth}p{0.62\linewidth}}
\toprule
\textbf{Constraint} & \textbf{Design decision} \\
\midrule
Zero egress (C1) & All models local via Ollama; synthetic training data
generated by a local 12B model; no cloud evaluation \\
Sensitive PII in tickets & Two-pass PII masking before indexing; aggregate
statistics only in this paper \\
Read-only mandate (C2) & Assistant drafts SQL with placeholders; rule-based
interceptors refuse live-data lookups; DB connector is SELECT-only on a
least-privilege account \\
8~GB VRAM (C3) & Quantized 4B/12B generators; embedder fine-tune with 8-bit
optimizer + gradient checkpointing; reranker disabled until production GPU \\
Live corpus (tickets arrive daily) & Incremental ingestion; per-class
metadata (\texttt{doc\_class}) and idempotent re-indexing \\
\bottomrule
\end{tabular}
\end{table}

\section{System Architecture}

\begin{figure}[htbp]
\centering
\includegraphics[width=\linewidth]{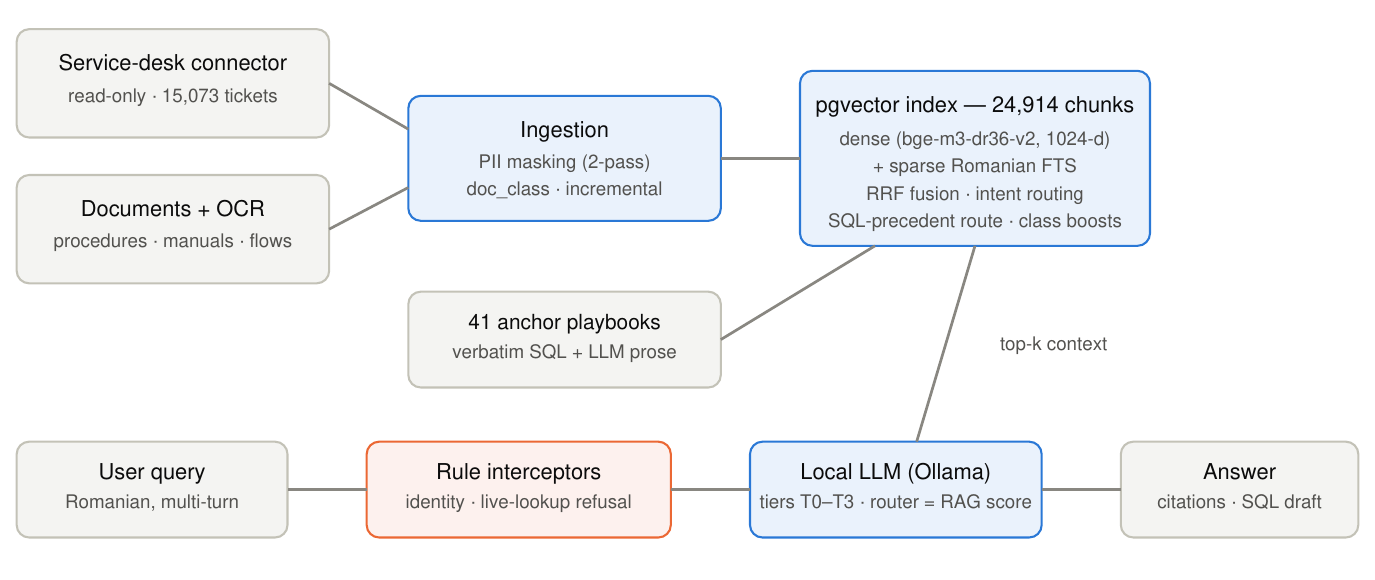}
\caption{RAGAL pipeline. Everything runs on-premise. The interceptors sit
\emph{before} the LLM: identity questions and live-data lookups are answered
deterministically and never reach the model. The 41 playbooks are distilled
offline from SQL-bearing ticket clusters (\S5.4) and indexed beside raw
tickets.}
\label{fig:pipeline}
\end{figure}

RAGAL is a conventional RAG stack assembled from open components ---
FastAPI, Streamlit, PostgreSQL with pgvector, Redis, Ollama --- with the
interesting decisions concentrated in retrieval and in guardrails
(Figure~\ref{fig:pipeline}):

\paragraph{Hybrid retrieval with intent routing.} Dense retrieval (bge-m3
\cite{chen2024m3}, 1024-d) is fused with sparse Romanian full-text search
via reciprocal rank fusion \cite{cormack2009rrf}. Lightweight intent
detection routes queries: procedure-type questions restrict and boost
documentary classes; database-modification questions activate a dedicated
candidate route restricted to chunks that \emph{contain} SQL, guaranteeing
that real precedents reach the context instead of being buried under
narrative tickets. On our internal golden evaluation, hybrid retrieval +
routing raised the score from 62\% to 81\% \emph{before any model was
fine-tuned} --- the cheapest points we bought anywhere in the project.

\paragraph{Escalation tiers.} T0 (routine, templated), T1 (single-shot RAG
--- the bulk), T2 (hard cases: workflow deadlocks, SQL drafting --- larger
local model), T3 (handoff to a human). The router signal is free: the
retrieval similarity score against past resolved tickets.

\paragraph{Rule-based interceptors.} Two classes of questions bypass the
LLM entirely: identity questions, and \emph{entity lookups} (``what is the
fiscal code of beneficiary X?'') which would otherwise tempt a small model
to present values from retrieved tickets as live data. The interceptor
replies deterministically that the assistant has no live database access,
and offers the read-only SELECT instead. On small models, we found
prompt-level discipline fragile and rule-level discipline reliable
(\S7.4).

\paragraph{Read-only enforcement in depth.} The assistant's DB connector
uses a least-privilege account (\texttt{db\_datareader} only), verified
empirically; a SELECT-only guard rejects any mutating statement as defense
in depth; and the product rule --- humans execute --- removes the remaining
risk.

\section{Corpus Construction}

\begin{figure}[htbp]
\centering
\includegraphics[width=\linewidth]{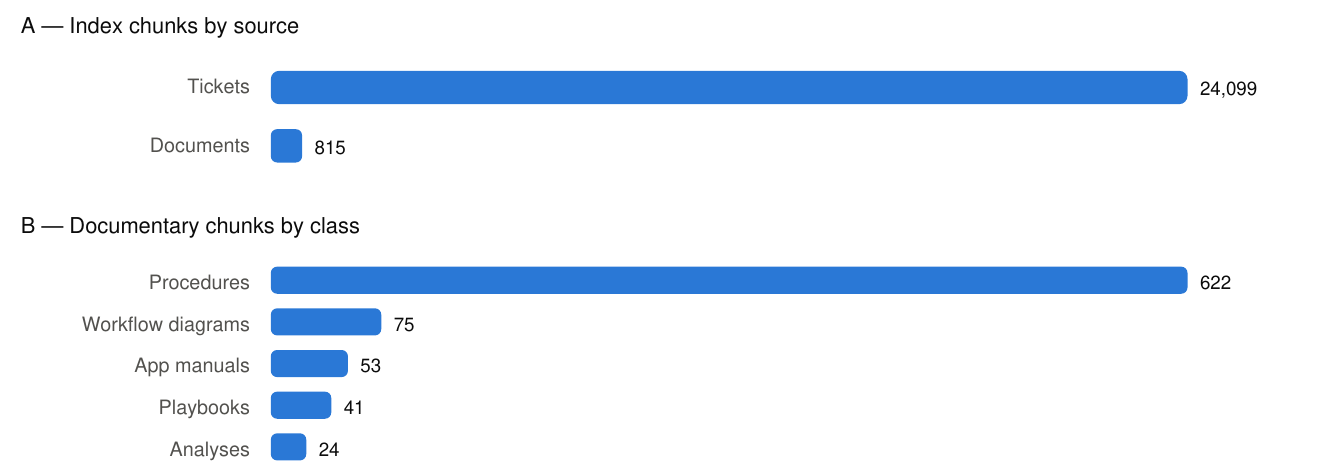}
\caption{Corpus composition (chunk counts; panels have independent scales).
Tickets dominate the index --- which is precisely why intent routing must
protect documentary classes from being buried (\S4), and why the
fine-tuning regression of \S6.3 stayed invisible without a per-domain
evaluation.}
\label{fig:corpus}
\end{figure}

The final index holds \textbf{24{,}914 chunks}: 24{,}099 from 15{,}073
resolved tickets, and 815 from documents (622 procedures, 75 workflow
diagrams, 53 application-manual chunks, 41 distilled playbooks, 24 analysis
documents); see Figure~\ref{fig:corpus}.

\subsection{Tickets: where the gold actually is}

\begin{figure}[htbp]
\centering
\includegraphics[width=0.85\linewidth]{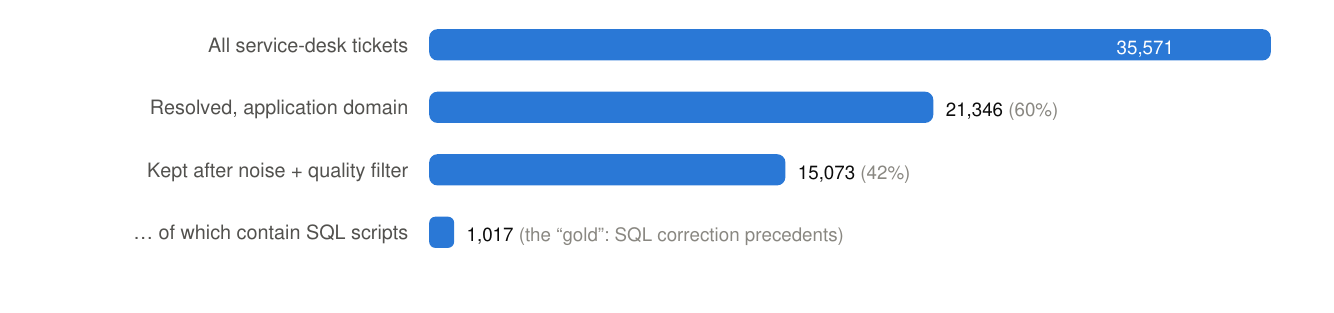}
\caption{The ticket funnel. Roughly 30\% of resolved application-domain
candidates are noise or rubber-stamp resolutions; the retained 15{,}073
tickets form the retrieval corpus and the fine-tuning pairs. The 1{,}017
SQL-bearing tickets --- most of whose scripts live in notes, not in the
resolution field --- are the highest-value slice and the source of the 41
playbooks (\S5.4).}
\label{fig:funnel}
\end{figure}

A read-only connector to the agency's service-desk platform paginates
resolved tickets in the application-support groups (21{,}346 candidates),
drops noise by subject/category (password resets, delegations), classifies
the resolution field (useful / workflow-log / rubber-stamp), and --- the
decisive discovery --- \textbf{fetches ticket notes}, because technicians
attach SQL correction scripts as notes on tickets whose resolution field
says merely ``solved''. Capturing notes raised SQL coverage from
${\sim}0.7\%$ to ${\sim}9\%$ of kept tickets (1{,}017 tickets with SQL).
Overall yield: 15{,}073 kept (${\sim}70\%$), each stored as \emph{problem
description $\rightarrow$ resolution + technical notes}
(Figure~\ref{fig:funnel}).

\subsection{PII masking --- hygiene that turned out to be a feature}

Two masking passes replaced identifiers with typed placeholders
(\texttt{<user>}, \texttt{<cui>}, \texttt{<cod\_proiect>}, \texttt{<nume>},
\texttt{<id>}): regex families for project codes, fiscal identifiers and
XML parameter dumps, plus a \textbf{dictionary of 417 technician usernames
harvested automatically} from structured fragments (XML tags, workflow-log
lines), filtered by document frequency to avoid masking common prose. The
second pass produced 64{,}701 maskings across 81\% of tickets, with zero
detected residuals on the target patterns and the SQL structure left
intact.

The unexpected result: after re-indexing the masked corpus, generation
\emph{improved} on our evaluation (\S7.3). With placeholders in context,
the model emits parameterized templates instead of copying a past
beneficiary's real values into the answer --- masking removed a copy-paste
attractor.

A third pass was triggered not by an audit but by \textbf{live
red-teaming}: a test question caused the assistant to reproduce a
technician's full name from a retrieved playbook. The username dictionary
did not cover full names written in free prose (e-mail headers,
signatures). We extended harvesting to \emph{structural} name contexts ---
the operator field of workflow-log lines and From/To/Cc e-mail headers ---
yielding a dictionary of 523 full names, masked in both token orders,
including names glued to adjacent words by table-cell concatenation. Two
safeguards proved essential. First, \textbf{dictionary validation against
false positives}: several local organizations are named after rivers and
villages whose names coincide with surnames; masking is applied only to
dictionary-confirmed person names, leaving toponymic organization names
intact. Second, a \textbf{document-frequency cap on \emph{all} harvested
candidates}: two common Romanian words (workflow-log field labels that
syntactically resembled usernames) had entered the strong dictionary and
were silently masked corpus-wide for eleven days --- degrading
${\sim}3{,}000$ chunks --- before a retrieved chunk exposed the artifact. A
username that appears in more than 1{,}000 of 15{,}073 tickets is prose,
not a person. The same pass added a rule for short quoted fiscal
identifiers in SQL literals (\texttt{SET CUI = '\ldots'}), below the length
threshold of the generic code rule. Repairs were applied \textbf{in place
with targeted re-embedding} of only the ${\sim}1{,}000$ affected chunks per
table (minutes, not a full re-index), followed by a full ticket re-ingest
once the dictionary was corrected.

\subsection{Documents, scans, and a warning about vision LLMs}

Documents are chunked structure-aware and tagged with a \texttt{doc\_class}
derived from their staging folder, enabling class-filtered retrieval and
incremental re-indexing. For scanned inputs (workflow diagrams, a scanned
ministerial order) we first tried a local vision LLM (gemma3:4b) as a
zero-install OCR; it \textbf{hallucinated plausible flowchart content that
did not exist in the image} --- disqualifying for normative text, where
fabricated-but-plausible is the worst failure mode. Plain tesseract
(Romanian traineddata, sparse page-segmentation for diagrams) produced
faithful, if unordered, label text that works well for retrieval. Each OCR
output carries a provenance header stating its origin and limitations.

\subsection{Playbooks by anchor distillation: hallucination-proof by construction}

The 1{,}017 SQL-bearing tickets cluster into recurring correction recipes.
A first, naive distillation (``synthesize a canonical playbook per cluster
with a local 14B model'') produced dangerous artifacts: on heterogeneous
clusters, 23 of 41 generated playbooks contained destructive statements
(DELETEs, auto-COMMIT) that no source ticket contained. We discarded them
and rebuilt with \textbf{anchor distillation}: pick the cluster's
\emph{medoid} ticket, copy its (masked) SQL \textbf{verbatim}, and let the
LLM write only the surrounding prose --- when to apply, symptoms, pitfalls,
verification steps. A splice guard strips any SQL the model tries to emit.
The result is a playbook whose executable content is by construction real;
the LLM contributes readability, not facts. The 41 playbooks are indexed
alongside raw tickets (dual RAG).

\section{Frugal Retriever Fine-Tuning}

\begin{figure}[htbp]
\centering
\includegraphics[width=0.9\linewidth]{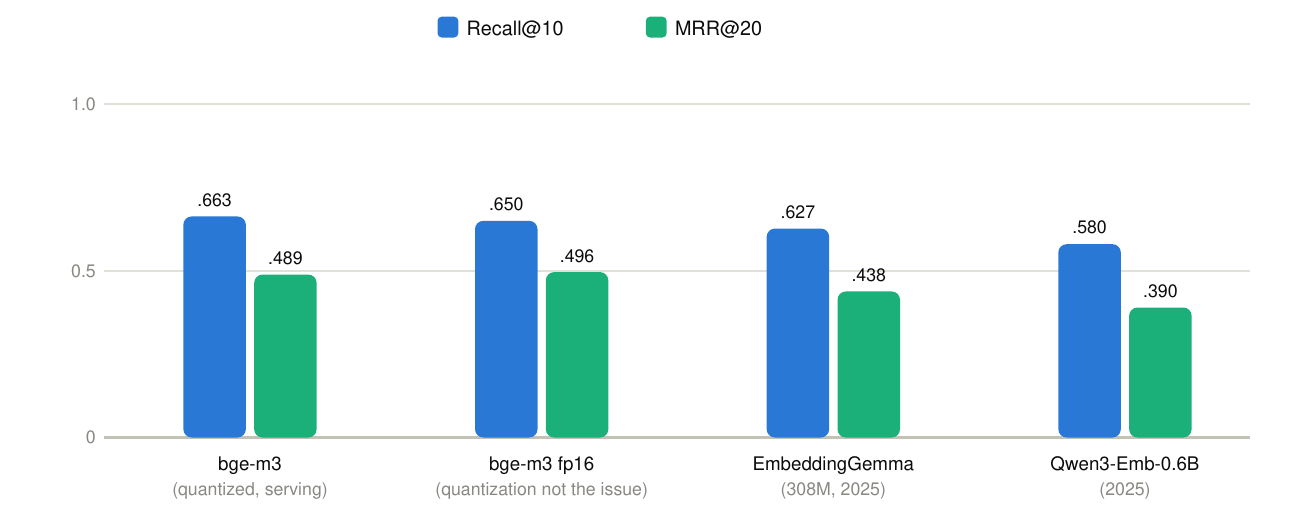}
\caption{Embedder bake-off on the ticket IR set (300 held-out queries /
4{,}000-ticket pool), before any fine-tuning. Neither of the two newer 2025
embedders beats the 2024 incumbent on our Romanian corpus, and the fp16
variant shows quantization costs nothing measurable --- evidence that
swap-the-model was the wrong lever and motivating the fine-tune of
\S6.1--6.4 instead. No public Romanian retrieval benchmark predicted this;
only local evaluation settled it (Lesson~8).}
\label{fig:bakeoff}
\end{figure}

Bake-offs on our own IR set (below) showed that newer or larger embedders
did not beat bge-m3 on this corpus --- Qwen3-Embedding-0.6B scored R@10
0.580 vs.\ bge-m3's 0.663, EmbeddingGemma 0.627, and an fp16
sentence-transformers bge-m3 matched the Ollama-quantized one (0.650 vs
0.663, within noise), ruling out quantization loss as a lever
(Figure~\ref{fig:bakeoff}). The remaining lever was fine-tuning bge-m3
itself on our data.

\subsection{Setup}

Training pairs mirror production traffic: \emph{query} = the user's raw
problem description (with inconsistent diacritics, as typed),
\emph{positive} = the masked subject + resolution of the same ticket.
4{,}835 pairs after deduplication and exclusion of the 300 IR-evaluation
hold-out tickets; 200 validation pairs. Loss:
CachedMultipleNegativesRanking (effective batch 64, mini-batch 8), lr 2e-5,
one epoch, max sequence 384, bf16. Two hard negatives per query were mined
with the current embedder. Diacritics are stripped at embedding time on
both sides (a stability fix: the quantized bge-m3 deterministically
produced NaN embeddings on certain Romanian diacritic sequences;
normalizing both corpus and queries into the same space fixed retrieval
consistency and the NaN crashes together).

\subsection{The 8~GB recipe --- and the silent spill that hides your OOM}

Full fine-tuning of a 568M model does not fit naively in 8~GB: AdamW in
fp32 needs roughly 9~GB for weights, gradients and moments alone. The
failure mode on Windows consumer hardware is treacherous: the NVIDIA driver
\textbf{silently spills to system memory instead of raising OOM}. Training
``works'', but the per-step time \emph{grows} ($375 \rightarrow 463
\rightarrow 525$~s/step in our first attempt) as tensors shuttle over PCIe.
\texttt{nvidia-smi} shows nothing unusual; the telltale sign is the per-process
\textbf{``GPU Process Memory / Shared Usage''} performance counter (we
observed 9.5~GB shared --- pure spill). Short smoke tests cannot detect
this: the degradation is progressive.

\begin{figure}[htbp]
\centering
\includegraphics[width=0.9\linewidth]{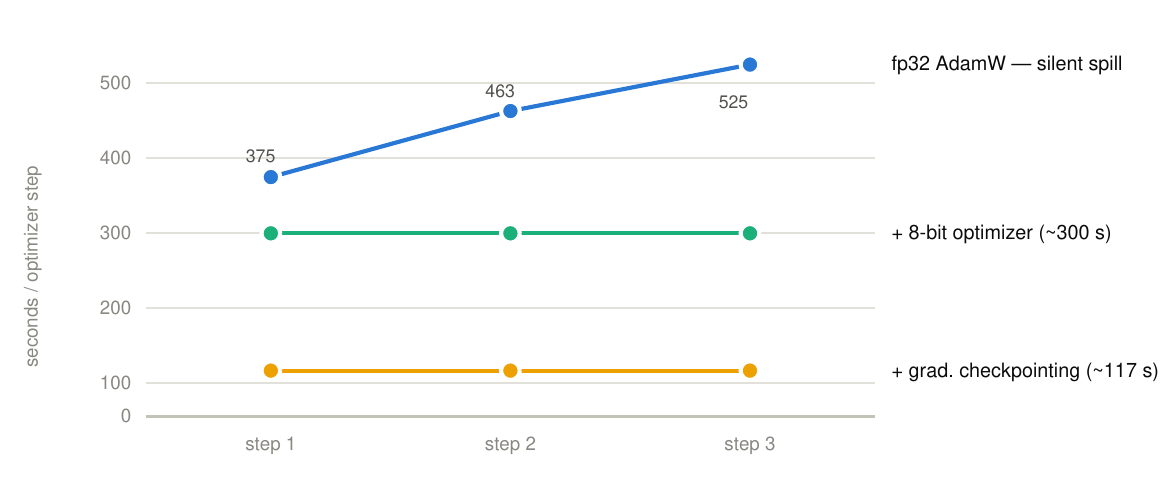}
\caption{Per-step training time on the 8~GB laptop under the three
configurations (representative measured steps, seq~384). With fp32 AdamW
the driver spills to system RAM silently and each step gets \emph{slower};
the 8-bit optimizer halts the growth; gradient checkpointing removes the
spill entirely --- a $3\times$ net win despite ${\sim}30\%$ extra compute.}
\label{fig:steptime}
\end{figure}

The fix that worked, in escalation order (Figure~\ref{fig:steptime}):
(1)~smaller mini-batch --- insufficient; (2)~8-bit optimizer
(\texttt{adamw\_bnb\_8bit}) \cite{dettmers2022bit} --- stopped the
degradation but plateaued at ${\sim}300$~s/step; (3)~\textbf{gradient
checkpointing} \cite{chen2016sublinear} (non-reentrant, compatible with the
cached loss) --- the winner: ${\sim}117$~s/step steady, no spill,
${\approx}2.5$~h/epoch, and 72 minutes for the second round (whose GenQ
pairs are shorter). Trading ${\sim}30\%$ extra compute to \emph{not} pay
PCIe round-trips was a $3\times$ net win. On a ${\geq}16$~GB GPU none of
this is necessary.

\subsection{Round 1: large gains, hidden damage}

\begin{table}[htbp]
\centering
\small
\caption{IR evaluation. Tickets: 300 held-out queries (raw user
descriptions) against a 4{,}000-ticket pool. Documents: 126 held-out
synthetic queries against an 815-chunk pool. ft-v1 = fine-tuned on tickets
only; ft-v2 = tickets + GenQ document pairs. $\downarrow$ marks regressions
below the stock baseline.}
\label{tab:ir}
\begin{tabular}{llccccc}
\toprule
\textbf{Eval set} & \textbf{Model} & \textbf{R@1} & \textbf{R@5} &
\textbf{R@10} & \textbf{R@20} & \textbf{MRR@20} \\
\midrule
Tickets & bge-m3 (stock) & 0.397 & 0.600 & 0.663 & 0.730 & 0.489 \\
Tickets & ft-v1 & \textbf{0.600} & \textbf{0.790} & \textbf{0.850} & 0.880 & \textbf{0.684} \\
Tickets & ft-v2 & 0.583 & 0.783 & \textbf{0.850} & \textbf{0.883} & 0.677 \\
\midrule
Documents & bge-m3 (stock) & 0.484 & 0.778 & \textbf{0.881} & 0.913 & 0.613 \\
Documents & ft-v1 & 0.421 & 0.722 & 0.849\,$\downarrow$ & 0.897 & 0.562\,$\downarrow$ \\
Documents & ft-v2 & \textbf{0.500} & \textbf{0.802} & \textbf{0.881} & \textbf{0.952} & \textbf{0.632} \\
\bottomrule
\end{tabular}
\end{table}

\begin{figure}[htbp]
\centering
\includegraphics[width=\linewidth]{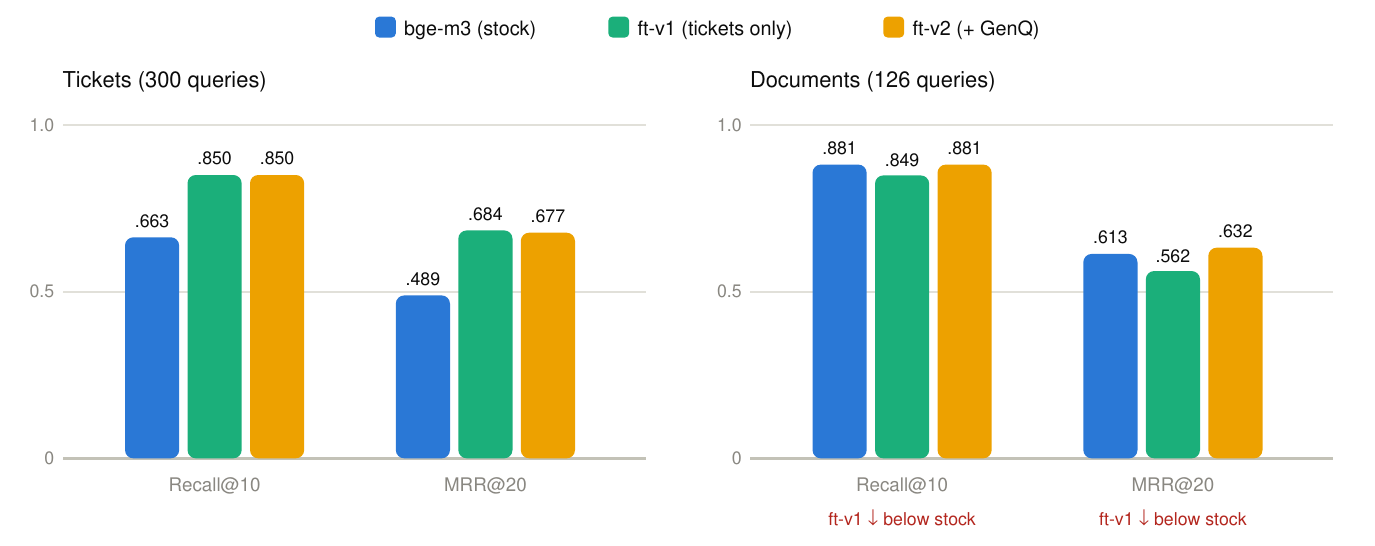}
\caption{The paper's central result. Left: fine-tuning on tickets lifts
ticket retrieval massively (R@10 $0.663 \rightarrow 0.850$) and the gain
survives round~2 intact. Right: the same round-1 model silently regressed
document retrieval below the stock baseline --- invisible without a
per-domain evaluation set; locally generated GenQ pairs (ft-v2) repair the
document domain to at-or-above stock on every metric while keeping the
ticket gains. Exact values in Table~\ref{tab:ir}.}
\label{fig:central}
\end{figure}

Round 1 (ft-v1) delivered the headline gain on tickets: recall@10 $0.663
\rightarrow 0.850$, +20 points at rank~1, MRR +40\% relative
(Table~\ref{tab:ir}, Figure~\ref{fig:central}). Validation loss decreased
then ticked up in the final third of the single epoch, arguing against
further epochs on 4.8k pairs.

But ft-v1 also did something we could not see at first: \textbf{it pushed
document retrieval below the stock model} (R@10 $0.881 \rightarrow 0.849$;
MRR $-5.1$ points). Training exclusively on ticket-style pairs deformed the
embedding space for the documentary domain. Nothing in the ticket
evaluation, the training curves, or casual use revealed this --- we found
it only after building a dedicated document evaluation set. We consider
this the paper's most generalizable warning: \textbf{when fine-tuning a
shared retriever on one domain of a multi-domain corpus, build per-domain
evaluations first; the damage is silent.}

\subsection{Round 2: zero-egress GenQ repair}

The repair had to respect C1: no external model may see the corpus. We
generated synthetic queries locally (GenQ): gemma3:12b via Ollama, JSON
output, three standalone Romanian questions per documentary chunk (815
chunks, ${\sim}10$~s/chunk, ${\sim}2.3$~h total, zero failures with an
incrementally resumable script). After deduplication: 1{,}962 (query,
chunk) pairs, plus a deterministic 15\% hold-out (126 queries) that became
the document evaluation set. Hard negatives were re-mined \emph{with ft-v1}
(the adapted model sees ``harder'' negatives), and training was repeated
once on the merged set (6{,}797 pairs; validation set unchanged for
comparability). Validation loss now decreased monotonically --- the larger,
more diverse set sustains a full epoch.

ft-v2 (Table~\ref{tab:ir}) kept the ticket gains (R@10 identical at 0.850;
R@1/MRR within noise on 300 queries) and \textbf{repaired the documentary
domain to at or above stock on every metric} (R@20 $0.913 \rightarrow
0.952$). ft-v2 is the serving model. One caveat, stated plainly: the
document queries are synthetic and in-distribution with the training style,
so absolute document numbers are optimistic; the model-to-model comparison
on the same set remains valid (\S9).

\section{End-to-End Evaluation}

\subsection{Methodology}

Besides the IR sets above, a \textbf{golden evaluation} of 21 real-workload
questions carries 27 machine-checkable assertions (retrieved classes,
citation presence, SQL presence/parameterization, refusal of live lookups,
admission of unknowns). A lesson that shaped the harness: even at
temperature 0 with a fixed seed, GPU inference is not bit-deterministic
(parallel reductions), so the harness separates \textbf{deterministic
assertions} (retrieval composition, interceptor behavior --- a hard
regression gate) from \textbf{generative assertions} (properties of LLM
output --- reported, but non-blocking). Without this split, the regression
gate flickers.

Following the masking incident of \S5.2, the harness gained an
\textbf{always-on anti-leak invariant}: every question implicitly asserts
that the answer contains no full person name from the PII dictionary (both
token orders). Because the corpus is masked, any name in an answer
indicates real contamination, so this invariant sits on the
\emph{deterministic}, blocking side of the gate.

\subsection{Serving the fine-tuned retriever}

ft-v2 is served through a switchable embedding backend
(sentence-transformers on GPU beside the default Ollama path), with the
re-embedded corpus in a parallel table for clean A/B against production.
Running the golden evaluation end-to-end on the ft-v2 pipeline (same 4B
generator as the baseline) passed the regression gate with \textbf{zero
deterministic regressions}: documentary-retrieval assertions 4/4, the
anti-contamination assertion and the live-lookup interceptor intact, and
\textbf{four generative assertions improved} over the stock-retriever
baseline --- parameterized-SQL drafting on two T2 correction cases and
unknown-admission on a discipline probe --- consistent with better
precedents reaching the context. Overall 20/27 assertions passed (74\%),
with the remaining failures in the band of known generative fluctuation of
the small evaluation generator.

\subsection{PII masking improved generation}

After masking pass 2 and re-indexing, the golden evaluation improved to
23/27 with zero deterministic regressions; the newly passing assertions
were precisely the T2 SQL-drafting ones (parameterized-SQL assertions on
the fiscal-code correction, workflow-resume and funding-note cases). The
mechanism is visible in transcripts: with masked precedents in context, the
model produces \texttt{UPDATE \ldots\ SET \ldots\ WHERE id = <id>}-style
templates instead of splicing a previous beneficiary's actual identifiers
into the draft.

\subsection{Model selection under a discipline criterion (negative results)}

Candidates for the T2 generator were judged on Romanian quality, schema
fidelity in drafted SQL, and above all \emph{discipline at the boundary of
knowledge}. gemma3:12b won our three-way live bake-off against qwen2.5:14b
(picked the correct table on 2/3 hard cases, admitted uncertainty) and
gemma3:4b (which once answered a Romanian SQL question in Korean --- small
models are T1-only). A community ``thinking/uncensored'' Qwen3.5-9B variant
was rejected after it \textbf{invented a ``1 million EUR'' legal ceiling}
under an explicit do-not-guess instruction and misdated the program period;
its English chain-of-thought also leaked into user-visible output. For
normative assistants, we recommend selecting on abstention behavior, not
generic benchmarks.

\subsection{A serving-configuration audit, and incident-driven hardening}

Late in the pilot, replaying \textbf{two real incidents from the live
queue} through the assistant exposed three defects that no synthetic
evaluation had caught --- and one confound that retroactively reframes the
numbers above.

\paragraph{Silent context truncation.} Our serving client never set the
runtime's context-window parameter, so the LLM server fell back to its
model default (${\sim}$4K tokens). Our RAG prompt --- ${\sim}$2K tokens of
behavioral rules plus top-k chunks --- routinely exceeded it, and the
runtime \textbf{truncates the head}: the rules (including ``answer in
Romanian'' and every guard) and the first-ranked documents were silently
cut. Symptoms we had misattributed to small-model weakness ---
self-introductions despite instructions, ignored guards, recitation of
late-context documents, and once an answer in the wrong language ---
disappeared after setting the window explicitly. All end-to-end scores in
\S7.2--7.3 were measured \emph{with} this defect present and are therefore
lower bounds. \textbf{Audit your serving defaults}: a runtime that
truncates silently converts a configuration omission into behavior that
imitates model failure.

\paragraph{Exact-precedent burial.} One incident had a near-verbatim
precedent in the corpus (the same templated complaint filed by a different
applicant, resolved weeks earlier). Dense retrieval ranked it \#1--2, yet
the answer never mentioned it: the query's procedural vocabulary triggered
the document-class boost, which pushed eight manual chunks above the
ticket. The fix is a \textbf{guard, not a weight}: the top dense-general
candidates are guaranteed a slot in the final context --- and are placed
\emph{first}, because with a mid-size generator, position in context
decides whether a precedent is used or merely present. After the fix, the
assistant answered by citing the precedent and its resolution.

\paragraph{Cross-operation script adaptation.} The other incident requested
a database change for which no precedent script existed; the model adapted
a script from a \emph{different} operation, inventing the target table name
in brackets and duplicating the UPDATE inside and outside the transaction.
We added an \textbf{operation guard} to the prompt contract: a precedent
script may be reused only for the same field/column it modifies; otherwise
the assistant must present the nearest precedents, state the limit
explicitly, and defer schema questions to the database team. Replaying the
incident after the fix produced exactly this behavior.

With all fixes in place, the corpus repairs of \S5.2, and 87 distilled
manual procedure sheets added to the index, the golden evaluation was
re-run as a fresh baseline: \textbf{48/48 assertions passed (100\%)} ---
all 27 original assertions plus the 21 always-on anti-leak invariants. The
trajectory ($74\% \rightarrow 85\% \rightarrow 100\%$ as retrieval,
masking, serving configuration and prompt guards were successively fixed)
supports the paper's central claim: in a constrained deployment, most of
what looks like model inadequacy is recoverable system engineering.

\subsection{Production serving: a 3-bit model that fits beats a 16-bit model that spills}

The production machine (one RTX 5060, 8~GB VRAM, 64~GB RAM) reproduced the
laptop's dilemma at deployment scale: the winning 12B generator does not
fit in VRAM at serving context sizes, and partial CPU offload through the
default runtime yielded ${\sim}20$~tok/s and 2--3-minute answers on hard T2
cases --- quality worth keeping, latency users would not accept. The
resolution was \textbf{aggressive quantization measured against the
evaluation gate, not against quantization folklore}: a Q3\_K\_M build
(5.7~GB) served by \texttt{llama-server} with flash attention and 8-bit KV
cache fits all layers on the GPU at a 16K context and delivers
\textbf{42~tok/s decode (${\sim}1{,}400$~tok/s prefill)} --- end-to-end
hard-case latency dropped from minutes to \textbf{12.6~s}. On the golden
evaluation the 3-bit model scored \textbf{48/48 --- strictly better than
the same model served in full precision with CPU spill}, which had suffered
citation misses and timeouts. Two serving traps found on the way
generalize: (1)~the runtime's chat template silently enabled
chain-of-thought, and on hard questions the model spent the entire token
budget ``thinking'', returning an \emph{empty answer with no error} --- an
eval collapse that imitated model failure until the reasoning budget was
pinned to zero; (2)~benchmarks taken while a previous runtime's zombie
process still held VRAM produced plausible but wrong baselines --- check
GPU process lists before trusting any number.

\subsection{A slow local judge: borrowing frontier-class judgment without egress}

Zero egress rules out the standard remedy for evaluating generative quality
--- a frontier cloud model as judge. Our workaround inverts the usual
speed/quality trade: a \textbf{744B-parameter MoE model (GLM-5.2, int4)
runs on the same production box, on CPU}, its dense core (${\sim}10$~GB)
resident in RAM and its 347~GB of experts streamed from NVMe through a
warm-expert cache sized by the available memory (26~GB pinned, ${\sim}31\%$
hit rate after cache warm-up). Decoding at 0.2--0.3~tok/s it is useless
interactively --- but \textbf{prefill scales sub-linearly} (a 708-token
judging prompt prefills in ${\sim}73$~s), so a compact verdict costs
${\sim}$10--13 minutes: an overnight or weekend budget of hundreds of
judgments from a model two orders of magnitude larger than anything the
hardware can serve interactively. The judge runs in batch against the live
assistant's answers with a resumable queue and emits structured verdicts
(agreement with a reference, hallucination flags, a corpus-gap flag). Its
first campaign judges the assistant against the \textbf{real resolutions of
closed support tickets} as ground truth, on a sample stratified by
application, testing a corpus-design hypothesis directly: whether tickets
from unrelated applications in the shared help-desk corpus dilute retrieval
precision for the target domain. A first retrieval-side result needed no
judge at all: the assistant surfaced the queried ticket itself in its top-3
sources for \textbf{95\% of the sample (100\% on the target domain)} --- in
this corpus, cross-application volume does not drown retrieval; whatever
quality gap exists must be sought in synthesis.

Three methodological lessons from the campaign's first day generalize to
anyone deploying an LLM judge. \textbf{(1)~Verdict formats must survive
truncation.} A pretty-printed JSON verdict that exceeds the token budget
dies silently and burns a 30-minute retry; demanding minified single-line
JSON with the \emph{scores first} and free-text last, plus a regex fallback
parser, made every verdict recoverable. \textbf{(2)~The rubric is not
neutral --- it encodes the system's role, and miscalibration swings like a
pendulum.} Our first rubric compared the assistant's answer to what the
human operator \emph{did}, and systematically punished the assistant for
\emph{proposing} rather than \emph{executing} --- a role mismatch with a
read-only-by-design system that manufactured a 60\% hallucination rate out
of thin air (an interim human audit of 20 verdicts caught it; the queue was
restarted). The corrected rubric eliminated the false flags but overshot
mildly in the other direction --- agreement concentrated on the middle
grade and one secondary field (the corpus-gap flag) degenerated to
always-true, teaching us to treat each verdict field as independently
falsifiable. \textbf{(3)~A judge two orders of magnitude larger is a second
opinion, not an oracle.} A human meta-audit of the first 21 corrected
verdicts found ${\sim}76\%$ sound, two false positives and one clean false
negative --- a case where the assistant's answer matched the ground-truth
resolution almost verbatim and the 744B judge failed it under both rubrics.
Our operating rule follows: per-application \emph{aggregates} drive
decisions; no individual verdict triggers a corpus edit without human
confirmation.

The full campaign completed over one weekend: \textbf{151 verdicts (130
stratified tickets + 21 golden questions), zero parse failures, 23.2 hours
of wall-clock judging (${\sim}9$ minutes per verdict)}. Aggregates: the
assistant's proposal was rated at least partially aligned with the real
resolution (\textbf{agreement ${\geq}1$) for 96\% of tickets}, mean quality
grade 4.17/5, with per-application means separating cleanly --- the target
domain's portal-facing application scored lowest (3.00) and the largest
ticket class second-lowest (3.84), against 4.3--4.6 elsewhere --- turning a
flat overall score into a ranked list of where corpus and synthesis work
should go next. The meta-audit extended to both extremes of the scale
generalized lesson~(3): of the five tickets the judge scored as outright
disagreement, human reading had already overturned three; low extreme
grades hid several near-verbatim reproductions of the real resolution,
while the rare top agreement scores were themselves inflated. Mid-scale
verdicts survived audit. A separate corpus forensics pass explained the
judge's most frequent substantive complaint (missing executable scripts):
with 1{,}024-character chunking, \textbf{73\% of the 1{,}362 tickets
containing executable SQL have the script entirely separated from the
problem-statement chunk} that anchors retrieval. The campaign's practical
output is thus threefold: a ranked per-application worklist, a
human-filtered shortlist of corpus additions, and one architectural fix ---
parent-document expansion at context assembly time (the retrieved ticket
chunk is replaced by its full 1--5~KB parent ticket), which leaves the
index, the fine-tuned embedder and the frozen evaluation gate untouched.
Deployed the same weekend, the fix passed the full 48-assertion gate with
zero regressions, and on re-asking the 27 SQL-bearing tickets of the judged
sample, coverage of the ground-truth resolution's key identifiers
(schema-qualified tables, stored procedures) in the assistant's answer rose
from \textbf{46\% to 83\%}, with complete transactional skeletons rising
from 19/27 to 23/27 --- measured by exact string matching, independent of
the judge. An overnight re-judging of the same 27 answers by the slow judge
showed no significant movement in its own metrics (agreement ${\geq}1$:
$27 \rightarrow 23$; mean grade $4.63 \rightarrow 4.41$; sign test
$p \approx 0.5$, on a class already at agreement saturation before the fix)
--- and its apparent regressions did not survive human reading (the two
unexplained downgrades were near-verbatim reproductions of the real
resolution, scored 0/1 by the judge), consistent with lesson~(3). We
therefore report the fix on the exact-match metric, with the judge as a
non-contradicting second opinion: zero new hallucination flags, saturation
maintained.

\section{Lessons Learned}

\begin{enumerate}
\item \textbf{Retrieval first, generator last.} Hybrid search + intent
routing: $62\% \rightarrow 81\%$ before any training; retriever fine-tune:
+19 points recall@10; the generator was upgraded only at the end.
\item \textbf{Single-domain fine-tuning silently regresses sibling
domains.} Build per-domain evaluations \emph{before} fine-tuning a shared
embedder.
\item \textbf{PII masking is a quality feature, not only compliance.}
Placeholders in context teach the model to emit templates instead of
copying values.
\item \textbf{Treat hallucination structurally, not rhetorically.} On
small models, prompts are fragile; interceptors and verbatim-anchored
distillation are reliable because they do not depend on the model's
obedience.
\item \textbf{Small vision LLMs are not OCR.} They fabricate plausible
content; tesseract remains the frugal standard for scanned normative text.
\item \textbf{Zero egress is fully workable}, including synthetic data: a
local 12B model generated the 1{,}962 GenQ pairs in ${\sim}2.5$ hours.
\item \textbf{On consumer GPUs, watch for silent VRAM spill, not OOM.}
Growing step-times are the symptom; per-process shared-memory counters are
the diagnostic; 8-bit optimizer + gradient checkpointing is the cure.
\item \textbf{New/hyped models must beat the incumbent on \emph{your}
corpus.} Three newer embedders and one community LLM variant all failed to
do so.
\item \textbf{Serving defaults are part of your model.} An unset
context-window parameter silently truncated the head of every long prompt
for weeks, producing failures indistinguishable from model weakness
(\S7.5). Config audits belong in the evaluation loop, not in incident
response.
\item \textbf{Replay real incidents; each one pays.} Every live incident we
replayed through the assistant surfaced a distinct, fixable defect --- a
PII residual class, a routing bias, a fabrication mode --- that 21 curated
questions had not. Incident replay is the cheapest red-teaming available to
an internal team, and each fix hardened into a permanent guard or a
blocking assertion.
\item \textbf{Quantize to fit, and judge the quant with your own gate.} A
3-bit build serving entirely from VRAM beat the same model in full
precision with CPU spill --- on quality (48/48 vs.\ citation misses and
timeouts), not just speed (\S7.6). Perplexity folklore would have predicted
the opposite; the evaluation gate settled it in an afternoon.
\item \textbf{Harvested dictionaries need frequency caps.}
Structured-context harvesting is precise but not immune: log field labels
that look like usernames poisoned our PII dictionary and silently degraded
${\sim}3{,}000$ chunks. Cap document frequency for \emph{every} candidate
source, however ``strong'' the context.
\item \textbf{A judge two orders of magnitude larger is a second opinion,
not an oracle.} Slow batch judging is affordable even on frugal hardware and
it earned its keep --- it produced a per-application worklist and the lead
that became our chunking fix --- but a human meta-audit found ${\sim}76\%$ of
verdicts sound, with both extremes of its scale unreliable. Use aggregates to
decide where to look; require human reading before acting on any individual
verdict (\S7.7).
\end{enumerate}

\section{Limitations}

The document-domain evaluation uses synthetic (GenQ) queries that share the
generation style of the training pairs; absolute numbers on documents are
in-distribution and optimistic, though same-set model comparisons remain
informative. All results come from one corpus, one language (Romanian) and
one institution; IR test sets are modest (300 and 126 queries). The pilot
lacks a formal user study --- feedback is collected in-app and will inform
future training rounds. The generator is used off-the-shelf; generator
fine-tuning (e.g., Finetune-RAG-style factual discipline) is future work.
The production GPU deployment is now live and reflected in \S7.6; the
slow-judge campaign of \S7.7 is complete, but its verdict-level reliability
analysis rests on a modest human audit (45 of 151 verdicts) concentrated
deliberately on the extremes of the scale.

\section{Conclusion}

A small team with no budget, no cluster and no permission to use the cloud
built a useful, safe, Romanian-language RAG assistant for government
technical support --- by spending effort where measurement said it
mattered: corpus quality (notes, masking, playbooks), retrieval
engineering, and a 72-minute embedder fine-tune on a laptop. We offer our
constraint-to-decision map, our 8~GB training recipe, and our
silent-regression warning to the many institutions whose data cannot leave
the building. Resource constraints did not prevent the system from working;
they forced the discipline that made it work.

\section*{Acknowledgments}

We thank our colleagues at AFIR's IT directorate for centralizing the
document corpus and testing the assistant, and the agency's management for
approving this publication.

\section*{Reproducibility statement}

Aggregate statistics only; no ticket text, personal data, internal schema
names or infrastructure details appear in this paper. Example SQL fragments
are structurally equivalent fabrications. Models used: bge-m3 (MIT), gemma3
4B/12B (Gemma Terms of Use), served via Ollama; fine-tuning with
sentence-transformers, bitsandbytes. Sanitized generic scripts (mining,
training, evaluation) are available at
\url{https://github.com/danmusetoiu/RAGAL}.

\end{document}